\documentstyle[aps]{revtex}
\begin{document}
\bibliographystyle{prsty}
\baselineskip=5.5mm
\parindent=7mm
\begin{center}
{\large {\bf \sc {Calculation of the quark condensate through Schwinger-Dyson equation }}} \\[2mm]
Zhi-Gang Wang $^{1}$ \footnote{E-mail,wangzg@mail.ihep.ac.cn}, Shao-Long Wan  $^{2}$ \\
 $^{1}$Institute of High Energy Physics, P.O.Box 918 ,Beijing 100039,P. R. China \\
$^{2}$Department of Astronomy and Applied Physics, University of Science and \\
Technology of China, Hefei 230026

\end{center}

\begin{abstract}
 In this letter, we clarify the algebra expression for calculating
 the quark condensate based on the non-perturbative quark propagator
 calculated through Schwinger-Dyson equation. The quark condensates,
 which characterize the low energy QCD
 vacuum, should not get a divergent quantity at large energy
 scale; the re-normalization group evolution behaviour at large
 energy scale therefore should be interpreted as "smeared collective effects" for
 it contains both perturbative and non-perturbative parts.
 We prefer the integral expression and get a quantity which is both
 convergent and scale dependent.
\end{abstract}



\section{Introduction}

Quark condensates, the non-vanishing vacuum expectation  values of
normal product  $\bar{q}q$, which characterize  the non-perturbative
quantum chromodynamics (QCD) vacuum, correspond to the breaking of
Chiral invariance. Their phenomenological importance to hardronic
physics properties exhibits itself by the initial  work of
 Shifman, Vainshtein and Zakharov \cite{Shifman}, which is
now known as QCD sum rule. The QCD sum rule approach tries to
incorporate non-perturbative elements of full
 QCD. As the starting point, operator product expansion  method
 was used to expand the time ordered currents into a series of quark and
 gluon condensates which parameterize the long distance properties, while
 the short distance effects are incorporated in the Wilson coefficients.
 However, these elements (condensates) can not yet be rigorous in
 QCD. In this letter, we clarify the algebra expression for calculating
 the quark condensate based on the non-perturbative quark propagator
 calculated through Schwinger-Dyson equation. The letter is arranged as:
 in section 2, re-normalization  group evolution of the quark
 condensate; in section 3, calculation of the  quark propagator and
 algebra expression for the quark condensate; in section 4,
 conclusion.

\section{Re-normalization  group evolution of the quark condensate}

Let us begin our work by the Gell-Mann-Oakes-Renner (GMOR)
relation \cite{Gell}, which announces the
existence of non-zero quark condensate  for the first time.

\begin{equation}
f^2_{\pi}m^2_{\pi}=(m_{u}+m_{d})_{\mu} \{ \langle \bar{u}u \rangle_{\mu} + \langle \bar{d}d
\rangle_{\mu} \}.
\end{equation}
This relation can be easily obtained by a few algebra
manipulation. The left hand side of Eq.(1) is scale independent,
while the components of the right hand side is scale dependent. To
make a scale independent quantity, the $m_{\mu}$ and $\langle \bar{q}q
\rangle_{\mu}$ must evolve in  opposite direction:
\begin{eqnarray}
m_{\mu}=\{ \frac{\log(\Lambda^2 / \Lambda^2_{QCD})}{\log(\mu^2 /
\Lambda^2_{QCD})}\}^{d} m_{\Lambda}, \ \ \ \langle \bar{q}q \rangle_{\mu}
=\{ \frac{\log(\mu^2 / \Lambda^2_{QCD})}{\log(\Lambda^2 /
\Lambda^2_{QCD})}\}^d \langle \bar{q}q \rangle_{\Lambda} =
\{ \log(\mu^2/ \Lambda^2_{QCD}) \}^d \  constant,
\end{eqnarray}
where $\Lambda_{QCD}$  is the QCD scale parameter , $\Lambda$
serves as boundary condition, $d$ is the anomaly dimension.
The $\langle \bar{q} q \rangle_{\mu}$, which characterizes the low
energy non-perturbative QCD vacuum, acquires a divergent quantity
at large energy scale, while the perturbative effects are
dominating. It is obvious an artifact of the re-normalization group equation.
We can borrow some idea from quantum electrodynamics (QED),
if we take QED as
self consistent theory,
to avoid the Landau singularity and triviality \cite{Bogolyubov},
 the perturbative expansion of coupling must break
down at some energy scale far below Landau singularity and
the QED  emerges into grand unified gauge theory which is
an asymptotic gauge theory, through not very successful \cite{Ross} .
It is an indication there must be some constrains for the
application of the re-normalization group equation.

Considering the particular scale behaviour of non-abelian gauge theory,
we can divide the QCD vacuum into both non-perturbative and
perturbative parts. The full two point quark Green function can
then be expressed into the following form:
\begin{eqnarray}
S(x)= S_{PT}(x)_{\mu}+S_{NP}(x)_{\mu}.
\end{eqnarray}
The scale $\mu$ is the Chiral phase transition point,
the effects above which can be
calculated by the usual perturbative expansion method,
while the effects below is embodied in the quark vacuum condensate.
 For example,
for $\mu > m_c$, the values of $\alpha_{s}(\mu)$
are sufficiently small that the effects of strong interaction can
be treated in perturbation theory.
 When one moves to low energy
scale, $\alpha_{s}$ increases. At $\mu \approx 1 GeV$ and
 $\Lambda^{(3)}_{\overline{MS}}$ (Here 3 is the number of quark flavor),
 $\alpha_{s}>0.5$. It signals the breakdown of perturbation expansion.
We can  also get support from renormalon
 theory of QCD, which
serves as a bridge between perturbative and non-perturbative QCD.
In general, ultraviolet renormalon can be canceled by introducing
counter terms while the infrared renormalon is absorbed into vacuum
condensate due to the strong coupling at low energy scale \cite{Beneke}.
 If the vacuum condensate  be pure collective non-perturbative objects,
the $\mu$ can not taken to be very large where the perturbative
effects are dominating.

\section{Calculation of the  quark propagator and
 algebra expression for the quark condensate }

If the full two point quark Green function is known, the
calculation of the quark condensate can be greatly facilitated. To
obtain the full quark Green function, let us first examine the
status of two main non-perturbative method, the Schwinger-Dyson(SD)
equation and lattice simulation.

Quantum chromodynamics, as a non-breaking $SU(3)$ gauge theory, merits
asymptotic freedom when the energy scale is large, which protects
the perturbative calculation doing the work, so we can get the full
two point quark Green function  by loops
calculation. However, at small energy scale,
the expansion of large coupling constant leads to a divergent
series. On the other hand, one of the main non-perturbative method,
lattice calculation can not give reliable results below 1 GeV,
where the most novel physics are supposed to lie. Furthermore,
lattice calculation  suffers its own shortcomings, such as Gribov
copies, boundary conditions, extrapolation and so on. The most
outstanding may be that the lattice simulation can not
include the contribution from the quark loops.
Neglecting the quark loops may lead
to very different results comparing with the corresponding full
ones. For example, although the ghost contributions are important
for the protection of unitary condition for the transition
amplitude, they can be safely neglected for its small numerical values
at high energy scales. However, at low energy scale, including ghost
greatly changes the infrared structure of  gluons two point
functions. That is, one get an infrared vanished instead an
infrared enhanced gluon
propagator \cite{Alkofer,Roberts} (Ref.\cite{Alkofer,Roberts} are
comprehensive review).

We can not solve the Euler-Lagrange equation for
quarks, gluons and ghost, and get the analytic expression for
those fields. At the present time SD equation is the most
reliable tool for studying
the infrared behaviour of the quarks and gluons propagators in the continuum limit,
 while it has its own shortcomings.
 One can get a series of  coupled re-normalized integral  equations
for those Green functions through functional integral method by including
the counter terms.
However, it is a divergent series and difficult to deal with in
practice, one have to make some truncation.
In fact, the long part of dressed vertex can be
approximated through the solution of the corresponding
Ward-Takahashi or Slavnov-Taylor identity \cite{Marciano} while
the transverse part of vertex can  be determined with the guide of
high energy behaviour, for example, Ball-Chiu vertex \cite{Ball} and
Curtis-Pennington vertex \cite{Curtis}.
The re-normalized SD equation for quark propagator can be expressed
in the following form:
\begin{equation}
S^{-1}(p,\mu)=Z_{2}(\mu,\Lambda)[\gamma \cdot p - m_{0}(\Lambda)]+\frac{16 \pi
i}{3}Z_{1}(\mu,\Lambda)\int^{\Lambda}
\frac {d^{4}k}{(2 \pi)^{4}} \gamma_{\mu}
S(k,\mu)\Gamma_{\nu}G^{\mu \nu}(k-p),
\end{equation}
where
\begin{eqnarray}
S^{-1}(p,\mu)&=& A(p,\mu)\gamma \cdot p-B(p,\mu)\equiv A(p,\mu)
[\gamma \cdot p-m(p)], \\
G^{\mu \nu }(k)&=&-(g^{\mu \nu}-\frac{k^{\mu}k^{\nu}}{k^2})G(k^2),
\end{eqnarray}
and $m_{0}$ stands for an explicit quark mass-breaking term.
With the explicit small mass term, we can preclude the zero
solution for $B(p)$ and in fact there indeed exist a bare current
quark mass which is given by the Yukawa coupling of Higgs due to the
spontaneous breaking of SU(2) Chiral symmetry at an energy scale
about 246 GeV. The full vertex $\Gamma_{\mu}$ can be approximated
by the Ball-Chiu vertex  and
Curtis-Pennington vertex.
This dressing comprises the notation of constituent quarks by
 providing a mass $m(p)=B(p,\mu)/A(p,\mu)$, which is corresponding to
 dynamical symmetry breaking. Because the form of
 the gluon propagator $g^2G(p)$ in the infrared region is unknown,
 one often one often uses model forms as input in the previous studies
  of the rainbow SD equation
  \cite{Tandy,Roberts}.
Although the loops integration is in general divergent,
the full quark gluon loop integration maybe non divergent.
There are some successful model dependent formulations for
the coupling constant  which insure
a natural integral cut off at large energy scale, for example,
confining $\delta^4(k)$ model, gauss distribution model, flat bottom model \cite{Tandy,Roberts,Wang}.
There is no need for the nomenclature
subtraction and hence re-normalization,
if there is no divergent in the loops integration. The
re-normalization coefficients can be set to 1.

The quark condensate, as isolated closed non-perturbative quark
 loop integral, if dependent on any energy scale, the scale is the
 momentum cut off in the quark loop integral,
\begin{eqnarray}
\langle\bar{q}(x)q(0)\rangle_{\mu}
&=&(-4N_{c})\int^{\mu}_{0}\frac{d^4 p}{(2\pi)^4}\frac{B(p^2)}
{p^2A^2(p^2)+B^2(p^2)}e^{ipx}.
\end{eqnarray}
At x=0 the expression for the local condensate $ \langle  \bar{ q}(0) q(0) \rangle $
is recovered,
\begin{equation}
\langle  \bar q(0) q(0) \rangle_{\mu} = -\frac{12}{16 \pi^2}
\int_{0}^{ \mu} d s s\frac {B(s)}{sA^2(s)+B^2(s)}.
\end{equation}
In fact, the energy scale $\mu$ is implicitly determined
by the effective gluon propagator, we can use above expression as the definition
for the quark condensate. However, presently, the non-perturbative technique can not prove
the relation
 \begin{equation}
\int_{0}^{ \mu} d s s \frac {B(s)}{sA^2(s)+B^2(s)}
\sim (\log(\mu^2/\Lambda^2_{QCD}))^d,
 \end{equation}
at low energy scale. Here $d$ is the anomalous dimension and takes
the value $d=\frac{12}{33-2n_{f}}$.

 Operator product expansion has proven that at large Euclidean momentum,
 the effective quark mass evolves  as
 \begin{equation}
 m(q^2)_{q^2 \rightarrow \infty} = \frac{c}{q^2} \{ \log(q^2 /
 \Lambda^2_{QCD})
 \}^{d-1} +m(\mu) \{
 \frac{\log(\mu^2 / \Lambda^2_{QCD})}{\log(q^2 /
 \Lambda^2_{QCD})}\}^{d},
 \end{equation}
here $c=-\frac{4 \pi d}{3} \frac{\langle \bar{q}q
\rangle}{[ \log(\mu^2/\Lambda^2_{QCD})]^d}$ is some constant
independent of $\mu$ \cite{Politzer,Gasser}, this implies that
$\langle \bar{q} q \rangle_{\mu} \sim
[\log(\mu^2/\Lambda^2_{QCD})]^d$, but not the definition of Eq.(8) .

If we take the limit $\mu \rightarrow \infty$ and
  assume ultraviolet
 dominating, the above relation $\langle \bar{q} q \rangle_{\mu} \sim
 (\log(\mu^2/\Lambda^2_{QCD}))^d$ is indeed the case, although the
 quark condensate is only defined at low energy scale.

When the energy scale $\mu$ increase,
one obtain an increasing quantity.
It is obvious a great mount of perturbative effects come in the
integral, then $\langle \bar{q}q \rangle_{\mu}$ contains both
perturbative and non-perturbative parts and can be best interpreted
as "smeared collective effects",
due to the falling attraction force can hardly bind the quarks
together to form vacuum condensate. There is another description of
quark confinement, the propagator of a colored state should have no
singularities on the real time-like $p^2$ axial,
the absence of Kallen-Lehmann spectral
representation precludes the existence of free quarks \cite{Roberts1,Bjorken}.
The interaction between dressed quarks may diminish
at separation beyond the characteristic length scale at which
point the quark Green's function approaches its vacuum values,
but the vacuum's repulsive force due to the absence of a mass pole in
the dressed quark propagator again compresses the quarks
together. However, there is
a long way to go before the repulsive quark-vacuum interaction can be
quantified.

Here we borrow some idea from studies  of strong coupling quantum
electrodynamics, which is often used as an
abelian model for QCD. In those studies, the effective coupling
constant is often taken as constant, and there does exist divergent
in the loops integral, so  re-normalization is necessary \cite{Williams}. We
have to change above algebra formulation for the quark condensate
through using re-normalized quark propagator,
\begin{equation}
\langle \bar{q}q\rangle_{\mu}
=-4N_{c}\int^{\infty}_{0}\frac{d^4 p}{(2\pi)^4}
\frac{m(p^2)}{A(p,\mu)(p^2+m^2(p))},
\end{equation}
where
\begin{equation}
A(p,\mu)|_{p^2= \mu^2}=1.
\end{equation}

The effects from hard gluonic radiative corrections to the
 quark propagator are connected to a possible change of the re-normalization
 scale $\mu$ at which the condensates are defined.
 Those effects are of minor
 importance for the non-perturbative effects in the low and medium energy regions,
 and should be neglected as it is perturbative contribution if we take quark condensate
 as pure collective objects.
In fact, the separation is hard to perform,

On the other hand, if we take the GMOR relation as the starting
point, one should not take the scale be very large due to the
artifact of the re-normalization group for $\langle \bar{q}q
\rangle_{\mu}$.
 the evolution of quark condensate with $\mu$ can be
determined from  GMOR relation and re-normalization group equation,
 \begin{eqnarray}
 \langle \bar{q}q\rangle_{\mu} &=&\{ \frac{\log(\mu^2/\Lambda^2_{QCD})}{\log(\Lambda^2/\Lambda^2_{QCD})}
 \}^{d}\langle \bar{q}q\rangle_{\Lambda}.
 \end{eqnarray}
In the following, we write the expressions more explicitly:
\begin{eqnarray}
 \langle \bar{q}q\rangle_{\mu} & =&-\{ \frac{\log(\mu^2/\Lambda^2_{QCD})}{\log(\Lambda^2/\Lambda^2_{QCD})}
 \}^{d} \frac{12}{16 \pi^4} \int_{0}^{\Lambda} d^4 p \frac{B(p^2)}{A(p^2)^2
 p^2+B(p^2)^2}
\end{eqnarray}
and
\begin{eqnarray}
\langle \bar{q}q\rangle_{\mu}& =&-\{ \frac{\log(\mu^2/\Lambda^2_{QCD})}{\log(\Lambda^2/\Lambda^2_{QCD})}
 \}^{d} \frac{12}{16 \pi^4}\int^{\infty}_{0} d^4 p
 \frac{m(p^2)}{A(p,\Lambda)(p^2+m^2(p))},
 \end{eqnarray}
here one  should not take $\Lambda$ a large value. Equation (15)
and Eq.(14) are expressions for quark condensate with and without
 hard gluonic radiative corrections respectively.

In Ref. \cite{Maris}, the authors got a scale dependent of
$\langle \bar{q}q \rangle_{\mu}$
through analyzing  the Ward-Takahashi identity and the Beth-Salpeter
equation with the tacit acceptance of the GMOR relation.
\begin{eqnarray}
 \langle \bar{q}q\rangle_{\mu} &=&\{ \frac{\log(\mu^2/\Lambda^2_{QCD})}{\log(\Lambda^2/\Lambda^2_{QCD})}
 \}^{d}\langle \bar{q}q\rangle_{\Lambda} , \nonumber \\
& =&\{ \frac{\log(\mu^2/\Lambda^2_{QCD})}{\log(\Lambda^2/\Lambda^2_{QCD})}
 \}^{d} (-4N_{c})\int^{\Lambda}_{0}\frac{d^4 p}{(2\pi)^4}
 \frac{m(p^2)}{A(p,\mu)(p^2+m^2(p))},
 \end{eqnarray}
In fact, although the expression has explicit scale dependence, it
does not satisfy the re-normalization group equation due to the
scale dependence of  $A(p,\mu)$.

\section{conclusion}
In this letter, we examine the algebra definition of quark
condensate. To obtain the quark condensate, we have to get
the non-perturbative quark propagator through quark SD equation.
If there is no divergent in the loops integration, re-normalization
is unnecessary. The quark condensate, as isolated closed
non-perturbative quark
 loop integral, if dependent on any energy scale, the scale is the
 momentum cut off in the quark loop integral. Otherwise, for a
 re-normalized quark propagator, the subtraction point provided
 a natural energy scale, we should use the definition Eq.(11).
 As the  critical point for perturbative-nonperturbative phase
 transition is about 1 GeV, one should not get a quantity at large
 energy scale $\langle \bar{q} q \rangle_{\Lambda}$ and the then
 evolve to lower energy scale through re-normalization group
 equation. To get explicitly scale dependent expressions for quark condensate,
 one can use Eq.(14) and Eq.(15) with  $\Lambda$ taken not far above 1 GeV.
 At large energy scale, the falling attraction force can
 hardly bind the quark
together to form vacuum condensate, meanwhile, a great mount of
perturbative effects come in the integral,
the quark condensate
$\langle \bar{q}q \rangle_{\Lambda}$ contains both
perturbative and non-perturbative parts and can be best interpreted
as "smeared collective effects". If one take the quark condensate as pure collective
objects, the $\mu$ should not taken to be
large, one can also get some support  from renormalon theory of QCD .
\begin{center}
Acknowledgment
\end{center}

One of the author (S.L.Wan) would thank Natural Science Foundation of China,
Grant No. 19875048 for financial support.


\begin{thebibliography}{99}
\bibitem{Shifman} M. A. Shifman, A. I. Vainshtein, and V. I. Zakharov,
Nucl. Phys. {\bf B 147 }, 385 (1979) 448
\bibitem{Gell} M. Gell-Mann, R.Oakes and B. Renner, Phys. Rev. 175, 2195 (1968)
\bibitem{Bogolyubov} N. N. Bogolyubov and D. V. Shirkov,
Introduction to the Theory of Quantized Fields, John Wiley (1980)
\bibitem{Beneke} M. Beneke, Phys.Rept. {\bf 317}, 1 (1999)
\bibitem{Ross} G. G. Ross, Grand Unified Theories,
Benjamin/Cummings, Menlo Park, California (1984)

\bibitem{Roberts1} C. D. Roberts, A. G. Williams and G. Krein,
Int. J. Mod. Phys. {\bf A} 7, 5607, (1992)
\bibitem{Alkofer} R. Alkofer and L. V. Smekal, hep-ph/0007355

 \bibitem{Roberts} C. D. Roberts and A. G. Williams, Prog. Part. Nucl. Phys.{\bf 33},
 477 (1994)

\bibitem{Tandy} P. C. Tandy, Prog. Part. Nucl. Phys. 39, 117
(1997)
\bibitem{Bjorken} J. D. Bjorken and S. D. Drell, Relativistic
Quantum Field, McGraw-Hill (1965)
\bibitem{Marciano} Marciano W and  Pagels H 1978  Phys. Rep. {\bf 36} 137
\bibitem{Ball} Ball J S and  Chiu T W  1980  Phys. Rev. D {\bf 22} 2542
\bibitem{Curtis} Curtis D C and  Pennington M R 1990   Phys. Rev. D{\bf 42} 4165 ;  1991 {\bf 44} 536
\bibitem{Wang}  K. L. Wang and S. L. Wan,  Phys. Rev. D
{\bf 47}, 2098 (1993); Z. G. Wang and S.L.Wan, hep-ph/0204158
\bibitem{Politzer} H. D. Politzer  Nucl. Phys. B {\bf 117} (1976) 397
\bibitem{Gasser} J. Gasser and H. Leutwyler, Phys. Rep. {\bf 87},
77 (1982); L. J. Reinders, H. Rubinstein and S. Yazaki,
Phys. Rep. {\bf 127}, 1 (1985)
\bibitem{Maris} P. Maris, C.D. Roberts and P.C. Tandy, Phys. Lett. {\bf B 420}, 267
(1998)
\bibitem{Williams} A. Kizilersu, A. W. Schreiber and A. G.
Williams, Phys.Lett. B {\bf 499}, 261 (2001); F. T. Hawes, T. Sizer and A.  G.
Williams, Phys.Rev. D{\bf 55}, 3866 (1997); F. T. Hawes, A. G. Williams and C. D.
Roberts, Phys.Rev. D{\bf 54}, 5361 (1996)
\end{thebibliography}
\end{document}